\documentclass{article}
\usepackage{ijcai22}

\usepackage{times}
\usepackage{soul}
\usepackage{url}
\usepackage[utf8]{inputenc}
\usepackage[small]{caption}
\usepackage{graphicx}
\usepackage{amsmath}
\usepackage{amsthm}
\usepackage{booktabs}
\usepackage{algorithm}
\usepackage{algorithmic}
\usepackage{amsmath,amsfonts}
\usepackage{algorithm,algorithmic}
\usepackage{multirow,multicol}
\usepackage{graphicx}
\usepackage{textcomp}
\usepackage{xcolor}
\usepackage{booktabs}
\usepackage{comment}
\usepackage{footmisc}
\usepackage{mathrsfs}
\usepackage{enumitem}
\usepackage{array}
\usepackage{float}
\usepackage{hyperref}
\usepackage{units}
\usepackage{subfigure}
\usepackage{pdfpages}
\usepackage{colortbl}
\usepackage{arydshln}
\usepackage{dblfloatfix}
\usepackage[flushleft]{threeparttable}

\title{MLP4Rec: A Pure MLP Architecture for Sequential Recommendations} 

\author{Muyang Li$^1$\and Xiangyu Zhao$^{2}$\thanks{Xiangyu Zhao is corresponding author.} \and Chuan Lyu$^3$\and Minghao Zhao$^4$ \and Runze Wu$^4$\and Ruocheng Guo$^5$ \\
\affiliations $^1$University of Auckland, $^2$City University of Hong Kong, $^3$Zhejiang University, \\$^4$Fuxi AI Lab, Netease, $^5$Bytedance AI Lab\\
\emails mli330@aucklanduni.ac.nz, xianzhao@cityu.edu.hk, chuan\_lyu@zju.edu.cn, 
\{zhaominghao,wurunze1\}@corp.netease.com, rguo.asu@gmail.com }

\begin{document}
\maketitle
\begin{abstract}
Self-attention models have achieved state-of-the-art performance in sequential recommender systems by capturing the sequential dependencies among user-item interactions. 
However, they rely on positional embeddings to retain the sequential information, which may break the semantics of item embeddings.
In addition, most existing works assume that such sequential dependencies exist solely in the item embeddings, but neglect their existence among the item features.
In this work, we propose a novel sequential recommender system (MLP4Rec) based on the recent advances of MLP-based architectures, which is naturally sensitive to the order of items in a sequence. 
To be specific, we develop a tri-directional fusion scheme to coherently capture sequential, cross-channel and cross-feature correlations.
Extensive experiments demonstrate the effectiveness of MLP4Rec over various representative baselines upon two benchmark datasets. The simple architecture of MLP4Rec also leads to the linear computational complexity as well as much fewer model parameters than existing self-attention methods.

\end{abstract}

\vspace{-3mm}
\section{Introduction}
\label{sec:intro}

Accurately modeling the chronological behavior of users is a critical area of research in recommender systems. The primary challenge is to capture the sequential pattern of user interests across multiple items, which is typically dynamic. To address this issue, Sequential Recommender Systems (SRS) were proposed and have garnered considerable interests from both academia and industry. While many endeavors have been put into this field, the newly emerged self-attention mechanism~\cite{vaswani2017attention} has gained a dominant position in SRS. Recent works show that self-attention based models can significantly outperform other models, and have achieved state-of-the-art (SOTA) performances in SRS~\cite{kang2018self,zhang2019feature,sun2019bert4rec}. 

Despite the success of self-attention in sequential recommendations, some limitations can potentially restrict its further development and practical applications. 
First, self-attention and its cognate methods are insensitive to the sequential order of the input items, and therefore relies on extra process such as adding positional embeddings to the input sequence to make the model aware of the information contained in the order of sequence.
However, existing self-attention methods, combining item sequence and positional embeddings from two heterogeneous data types, may interrupt the underlying semantics of item embeddings~\cite{zheng2021rethinking}.
Second, self-attention methods' computational complexity is quadratic to the length of input item sequence, which yields an unneglectable computational cost for large-scale recommender systems. 
Third, incorporating self-attention in recommender systems typically leads to huge amounts of model parameters, which result in difficulty in model optimization and the increased chance of over-fitting.

\begin{figure}[t]
\centering
\includegraphics[width=\linewidth]{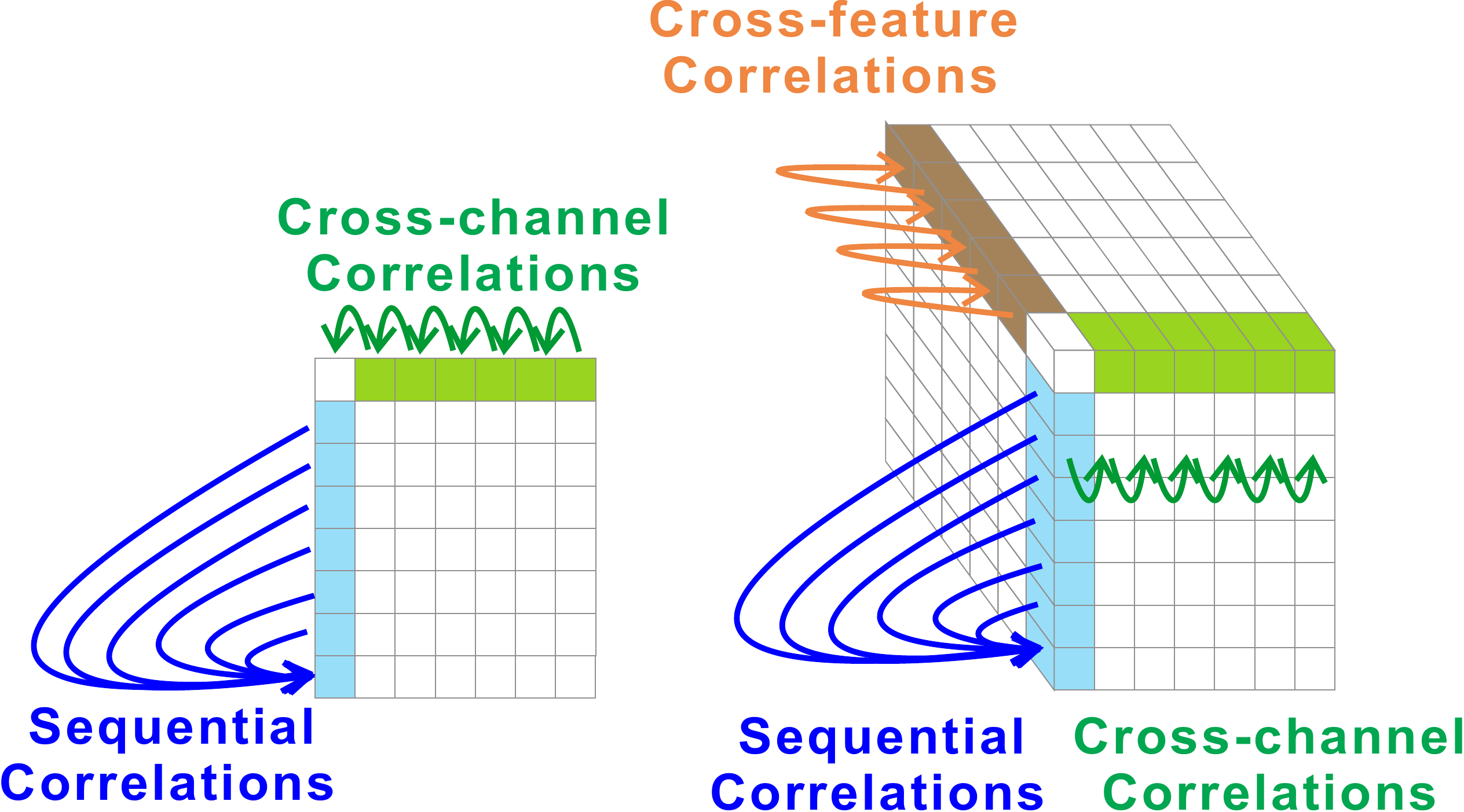}
\vspace{-5mm}
\caption{Bi-directional correlations v.s. Tri-directional correlations \label{fig:overflow}}
\vspace{-7mm}
\end{figure}

\begin{figure*}
    \centering
    \includegraphics[width=\linewidth]{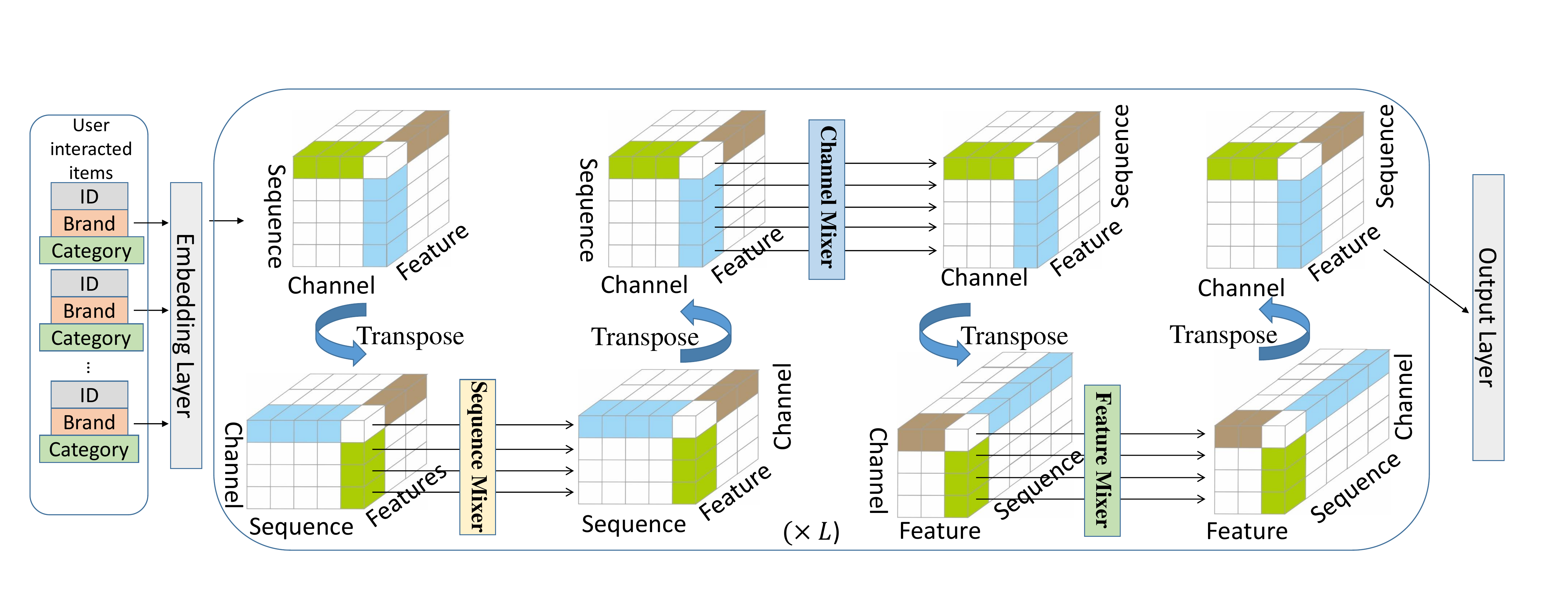}
    \vspace{-10mm}
    \caption{Overall framework of MLP4rec.}
    \label{fig:overview}
    \vspace{-5mm}
\end{figure*}

Recent advances in Multi-layer Perceptron (MLP) architectures, such as MLP-Mixer, gMLP and resMLP~\cite{tolstikhin2021mlp,liu2021pay,touvron2021resmlp}, show competitive performances in computer vision tasks despite their architectural simplicity and linear computational complexity. This questions the necessity of attention mechanisms and shows the possibility to replace them via simple MLP architectures. To address aforementioned challenges of self-attention based SRS, this paper proposes a simple yet effective \textbf{MLP} framework for sequential \textbf{Rec}ommendations (\textbf{MLP4Rec}), which has two-fold advantages. 
First, along with above MLP-based models, MLP4Rec is by design sensitive to the order of input item sequence, avoiding the bottleneck caused by using positional embeddings. 
Second, upon pure MLP blocks, MLP4Rec possesses linear computational complexity and significantly lower amount of model parameters than self-attention based SRS models.

However, due to the design of bi-directional mixer in existing MLP architectures~\cite{tolstikhin2021mlp,lee2021moi}, utilizing them for sequential recommendations can only capture the dependencies of item embeddings without taking the items' explicit features (e.g., brand and category) into consideration. 
To this end, we devise a novel tri-directional information fusion scheme for MLP4Rec with a cross-feature mixer, which enables the framework to capture the interactions among all item features (including ID feature), as illustrated in Figure~\ref{fig:overflow}. In addition, the tri-directional scheme also applies the classic bi-directional mixers from MLP-based models~\cite{tolstikhin2021mlp,lee2021moi} on item explicit features, which learns the users' sequential preferences within these features (e.g., loyal users of \textit{brand A} are likely to continue to buy \textit{brand A}'s products). 
Through extensive experiments, we demonstrate that MLP4Rec shows significantly superior performance than the state-of-the-art methods on two benchmark datasets. To summarize, this paper has the following contributions: 
\begin{itemize}[leftmargin=*] 
\vspace{-1mm}
\item We investigate the possibility of replacing self-attention mechanism  by simple MLP architectures for sequential recommendations;
\vspace{-1mm}
\item To the best of our knowledge, this is the first work that proposes a tri-directional mixing MLP architecture; 
\vspace{-1mm}
\item We validate the effectiveness of our proposed framework against SOTA models via extensive experiments on two benchmark datasets.
\end{itemize}

\section{Framework}
\label{sec:framework}
In this section, we discuss the framework, methodology and optimization of our proposed MLP4Rec framework. 

\subsection{Problem Formulation}
Follow commonly adapted settings \cite{li2018learning,kang2018self}, we denote the participant of interactions - users as \(U = \{u_1,...,u_n...,u_{N}\}\), where \(n\) indicates the ID of the user. Items as \(I = \{i_1,...,i_m...,i_{M}\}\), where \(m\) indicates the ID of the item. In addition, each item have some associated features, such as category and brand, we denote those features as \(Q = \{q_1^m,...,q_k^m,...q_K^m\}\), where \(q_k^m\) refers to the \(k\)-th feature of item \(m\). We sort the items that users have interacted with into sequences, thus each user has a corresponding sequence containing items (s)he once viewed chronologically. We denote the item sequence of user \(n\) as \(S_n = \{i_1,...,i_t...,i_{s}\}\), where \(i\) stands for item, \(t\) describes the chronological order of item, \(s\) is the maximum length of the sequence. 

With the notations and definitions above, the problem of next item recommendation can be formally defined as follows: \textit{Given user $u_n$'s historical item sequence $S_n$, the goal is to find a recommendation model $f$ to predict the next possible items $i_{s+1}$ for the given user, i.e., $f:S_n \rightarrow i_{s+1}$, which the target user is likely to interact with. }

\subsection{Framework Overview}

Here, we present our MLP-based SRS which can explicitly learn tri-directional information. As we mentioned before, in order to make an informed prediction, a model must be able to capture the 3-fold information. The first fold is, the temporal information i.e. sequential dependencies among \(S_n\). The second fold refers to, the interest information contained in the item embedding, since different channels (dimensions) of an item embedding represents different latent semantics, the cross-channel correlation is also important for our task. The third fold is the correlations among item features, collectively, they contribute to modeling the semantic meaning of an item.
By repetitively transposing and applying MLP blocks in different directions of the input embedding tensor as shown in Figure~\ref{fig:overview}, we show that our proposed framework can capture the sequential, cross-channel and cross-feature correlations at the same time. 

To be specific, MLP4Rec consists of \(L\) layers, where each layer has the identical setting, a sequence-mixer, a channel-mixer, and a feature-mixer. Following~\cite{tolstikhin2021mlp}, all \(L\) layers share the same parameters to reduce model parameters. Within each layer, we first apply independent sequence-mixers and channel-mixers for different features, so as to learn their unique characteristics. Then, we utilize a feature-mixer to learn correlations among all features.

\subsection{Detailed Architecture}
\textit{\textbf{Embedding Layer.}} We adapt a commonly used method for constructing item ID embeddings and feature embeddings, i.e., learning an embedding lookup table to project the discrete item identifiers (i.e., IDs) and explicit features (e.g., category and brand) into dense vector representations with dimension \(C\) \cite{cheng2016wide}. 
After the embedding layer, we can stack the embeddings of item IDs and explicit features into individual embedding tables, where the row of the embedding table is each embedding vector, the column of the embedding table contains channel information. Stacking all embedding tables together, we obtain a 3-d embedding table as shown in Figure 3. Note that, unlike self-attention models, our proposed model does not need to learn a positional embedding for an input sequence. Instead, temporal information can be directly learned via the sequence-mixer.

\begin{figure}[t]
    \centering
    \hspace*{-7mm}
    \includegraphics[width=1.16\linewidth]{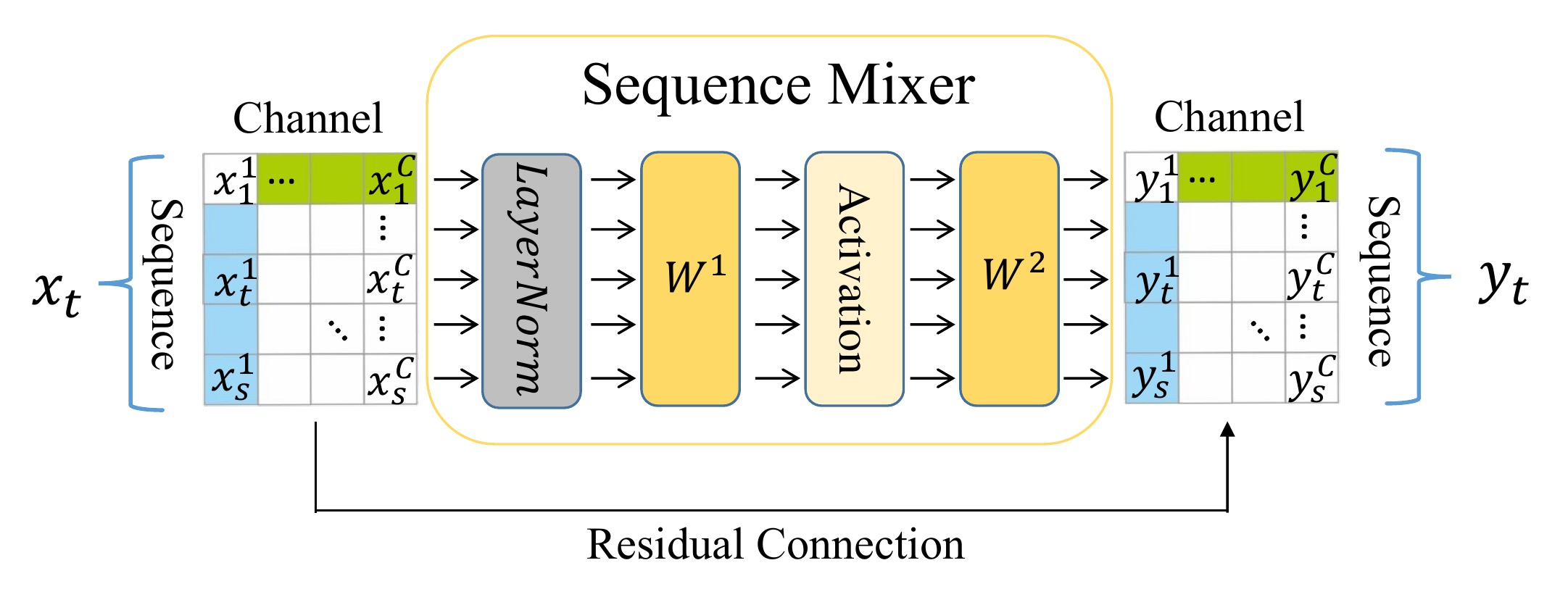}
    \vspace{-7mm}
    \caption{Architecture of Sequence-Mixer } 
    \label{fig:sequencemixer}
    \vspace{-5mm}
\end{figure}

\textit{\textbf{Sequence-mixer.}} The sequence-mixer is an MLP block, which aims to learn the sequential dependencies across the entire item sequence. The sequence-mixer block takes the rows of the embedding table as input features (applied to the transposed embedding table), and outputs an embedding table with the same dimension as the input. But in this output table all the sequential dependencies are fused within each output sequence. More specifically, a set of input feature would be the \(c\)-th dimension of each embedding vector across the whole sequence, i.e. \(\{{x_1^c,...,x_t^c,...,x_s^c}\}\) as shown in Figure ~\ref{fig:sequencemixer}. The correlation between them is sequential, which shows the evolvement of user interest across time, thus making sequence-mixer sensitive to the sequential order. Formally, we denote the output of sequence-mixer at layer \(l\) as:
\begin{equation} 
 \boldsymbol{y_t} = \boldsymbol{x_t} + \boldsymbol{W^2} g^l(\boldsymbol{W^1} LayerNorm(\boldsymbol{x_t}))
\end{equation}

where \(t\) = 1,..,\(s\). \(\boldsymbol{x_t}\) is the input feature, which is the embedding vector at time step \(t\). \(\boldsymbol{y_t}\) is the output of the block, \(g^l\) is the non-linear activation function at layer \(l\), \(\boldsymbol{W^1} \in \mathbb{R}^{r_s\times s}\) denotes the learnable weights representing the first fully connected layer in the sequence-mixer, \(\boldsymbol{W^2} \in \mathbb{R}^{s\times r_s}\) signifies the learnable weights of the second fully connected layer in the sequence-mixer, \(r_s\) is the tunable hidden size of sequence-mixer. We employ layer normalization (LayerNorm)~\cite{ba2016layer} and residual connection~\cite{he2016deep} as in MLP-mixer~\cite{tolstikhin2021mlp}.

\textit{\textbf{Channel-mixer.}} Like sequence-mixer, channel-mixer is also an MLP block with a similar macro architecture, their key distinction is between their purpose. The objective of channel-mixer is to learn the correlation within an embedding vector. The embedding of an item ID or item feature usually expresses some latent semantics on each dimension, learning their representation and internal correlation is crucial for recommendation. The channel-mixer takes the columns of the embedding table as input feature, as shown in Figure 2, channel-mixer is applied after transposing the embedding table back to its original shape. After the sequence-mixer, sequential information is fused within each sequence, but the cross-channel correlation has not been discovered yet. Channel-mixer will take \(t\)-th item embedding's dimension as input feature, i.e. \(\{{x_t^1,...,x_t^c,...,x_t^C}\}\), the correlation between them is cross-channel, collectively they express the overall semantic of the embedding. So after the channel-mixer, the cross-channel correlation will be fused within the output sequence. We can denote the output of channel-mixer at layer \(l\) as:
\begin{equation} 
 \boldsymbol{y_c} = \boldsymbol{x_c} + \boldsymbol{W^4} g^l(\boldsymbol{W^3} LayerNorm(\boldsymbol{x_c}))
\end{equation}
where \(c\) = 1,2,..,\(C\), \(\boldsymbol{x_c}\) is the input feature, which is the \(c\)th dimension across all embedding at time step \(t\), and \(\boldsymbol{y_c}\) is the output of the block,\(\boldsymbol{W^3} \in \mathbb{R}^{r_C\times C}\) is learnable weights of the first fully connected layer in the channel-mixer, \(\boldsymbol{W^4} \in \mathbb{R}^{C\times r_C}\) is learnable weights of the second fully connected layer, \(r_C\) is tunable hidden size in channel-mixer. 

\textit{\textbf{Feature-mixer.}} Now we present the feature-mixer, the key component to connect features together. After the sequence-mixer and the channel-mixer, the sequential and cross-channel dependencies are fused within each sequence. However, the information among the embedding table of different features is still independent of each other. The feature-mixer can fuse cross-feature correlation into the representation of each sequence. More importantly, since feature-mixer is the last block in a layer, which not only communicates feature information, but also shares the sequential and cross-channel dependencies within each feature to other features, thus coherently connects the tri-directional information. The feature-mixer acts on features dimension as shown in Figure ~\ref{fig:overview}. We denote the output of feature-mixer at layer \(l\) as:

\begin{equation} 
 \boldsymbol{y_k} = \boldsymbol{x_k} + \boldsymbol{W^6} g^l(\boldsymbol{W^5} LayerNorm(\boldsymbol{x_k}))
\end{equation}

where \(k\) = 1,2,...,\(K\), \(\boldsymbol{x_k}\) is the input feature, which is the embedding vector of \(k\)th feature at time step \(t\), and \(\boldsymbol{y_k}\) is the output of the block, \(\boldsymbol{W^5} \in \mathbb{R}^{r_K\times K}\) denotes the learnable weights of the first fully connected layer in the feature-mixer, \(\boldsymbol{W^6} \in \mathbb{R}^{K\times r_K}\) is the learnable weights of the second fully connected layer in the feature-mixer, and \(r_K\) is tunable hidden size in feature-mixer. 

\subsection{Training and Inference} 
\textbf{\textit{Training.}} Our training procedure follows the commonly used paradigm in SRS ~\cite{kang2018self,zhang2019feature}, using binary Cross-Entropy loss: 
\begin{equation} 
 L = -\sum_{S_n \in S}\sum_{t \in [1,...,s]}[log(\sigma({r_{i_t,t}})) + \sum_{j \not\in S_n}log(1-\sigma(r_{i_j,t}))]
\end{equation}
where \(\sigma\) demotes sigmoid function, \({r_{i_t,t}}\) is model's predicted similarity to ground-truth item \(i_t\), and \(r_{i_j,t}\) is the predicted similarity to sampled items at timestamp \(t\), \(j\) is the negative sampled items, \(S\) is the super set of all users' interaction sequences. Please find the detailed optimization algorithm in \textit{Section A of Technical Appendix}.

\textbf{\textit{Inference.}} We adapt the most commonly used inference method in SRS for fair comparison~\cite{kang2018self,zhang2019feature}. To be specific, after \(L\) layers of sequence-mixer, channel-mixer and feature-mixer, we obtain a sequence of hidden states that contains the sequential, cross-channel and cross-feature dependencies of each interaction, respectively. Assuming at time step \(t\), we wish to predict next item \(i_{t+1}\), given sequence of hidden states \(H = {{h_1,...,h_t}}\), we can calculate the cosine similarity between \(h_t\) and all candidates items \(E_m\) via dot product as:
\begin{equation} 
    r_{m,t} = h_t \cdot E_m^T
\end{equation}
where \(m\) = 1,...,\(M\), \(E_m \in \mathbb{R}^{M\times C}\) is the item embedding of all candidate items and \(r_{t,m}\) indicates the similarity between hidden state \(t\) to all candidate items, best \(K\) predictions will be ranked by their similarity. 

\subsection{Discussion}

\textbf{\textit{Relation to MLP-Mixer and resMLP}}. The key architectural differences of MLP4Rec to MLP-Mixer and resMLP is that MLP-Mixer and resMLP directly project a 3-dimensional input (image) into a 2-dimensional embedding table, and then operate 2-dimensional (spatial/channel) information fusion, whereas MLP4Rec directly operates on the 3-dimensional input and conducts the sequential/channel/feature information fusion. MLP4Rec can degenerate into MLP-Mixer and resMLP when input is a 2-dimensional embedding table and removing feature-mixer. 

\textbf{\textit{Complexity Analysis}}. The following discussion regarding to the time and space complexity of our model is for inference stage. (1) Time Complexity: MLP4Rec's time complexity is $O(s + C + K)$, which is linear complexity to the sequence length \(s\), embedding size \(C\) and feature number \(K\). Compared to the time complexity of self-attention, $O(s^{2}C + C^{2}s)$, the theoretical lower bound of the MLP4Rec's time complexity is significantly lower. (2) Space Complexity: MLP4Rec's space complexity is $O(K(s + C + 1))$, where the number of features \(K\) is usually limited, especially after feature selection. On the other hand, the space complexity of self-attention is $O(sC + C^2)$~\cite{kang2018self}, which is quadratic to the embedding size. In experiment part, we show that not only we keep a theoretical lower bound in space complexity, but in practice, we also achieved significantly lower number of parameters. Please find the detailed time and space complexity analysis in the \textit{Section B of Technical Appendix}.

\section{Experiments}
\label{sec:experiment}
This section evaluates the performance of MLP4Rec against representative baselines on two benchmark datasets.

\subsection{Datasets}
We choose two widely used datasets to benchmark our performance on both small and large datasets, and their statistics can be found in Table~\ref{fig:statistics}. (1) \textbf{MovieLens}\footnote{https://grouplens.org/datasets/movielens/100k/}: MovieLens is a site for recommending movies to users given their historical ratings, which is now one of the most commonly used benchmarks across the field of recommender system. We use MovieLens-100k in our experiments. 
(2) \textbf{Amazon Beauty}\footnote{http://jmcauley.ucsd.edu/data/amazon/}: The online reviews and ratings of Amazon are commonly used benchmarks for evaluating recommendations. We use the ``Beauty" category in our experiments. We filter out the items and users that have less than 5 interactions for two datasets. We set the maximum sequence length as 50 for both datasets, and conduct zero-padding for shorter sequences. The detailed item features in above datasets can be found in \textit{Section C of Technical Appendix}. 

\begin{table}[t]
\centering
\caption{Statistics of the datasets.}
\vspace{-3mm}
\begin{tabular}{c||cc} 
\hline
Data & MovieLens & Beauty  \\ 
\hline
\# interactions & 100,000 & 2,023,070  \\
\# users & 943 & 1,210,271  \\
\# items & 1,682 & 249,274 \\
\# avg. length & 106 &  8.8 \\
\hline
\end{tabular}
\vspace{-3mm}
\label{fig:statistics}
\end{table}

\subsection{Evaluation Settings}

We employ the commonly used evaluation method in SRS, namely next-item prediction. For dataset splitting, next-item prediction task uses the last item in an interaction sequence as the test set, the item before as validation set, and the rest of the items will be used as training set. Following common settings, we pair 100 negative samples with ground-truth items during prediction~\cite{kang2018self}.

\textbf{\textit{Metrics.}} We apply three commonly used evaluation metrics in recommender system, namely hit ratio (HR), normalized discounted cumulative gain (NDCG) and Mean Reciprocal Rank (MRR). HR measures the probability of the ground-truth item that appears in model's top-K recommendation, NDCG measures the order of the top-K recommended items generated by the recommender, and finally MRR takes the reciprocal of the ground-truth item's ranking in top-K recommendation and averages across all the evaluated items. All results are averaged on three random seeds. 

\begin{table*}
\centering
\caption{Overall performance comparison on two datasets, where best baseline performances are underlined}
\vspace{-3mm}
\scalebox{1.5}
\ADLnullwidehline
\arrayrulecolor{black}
\renewcommand{\arraystretch}{1.2}
\begin{tabular}{c||ccc||ccc} 
\hline
\multicolumn{1}{c||}{Methods} & \multicolumn{3}{c||}{MovieLens} & \multicolumn{3}{c}{Beauty}  \\ 

\multicolumn{1}{c||}{Metrics} & MRR@10 & NDCG@10 & HR@10 & MRR@10 & NDCG@10 & HR@10    \\
\hline
\multicolumn{1}{c||}{PopRec}                         & 0.1496 & 0.0783 &  0.4044       & 0.0305 & 0.0443 & 0.0954    \\
\multicolumn{1}{c||}{BPR}                             &    0.1479    &   0.1905     &  0.3474            &     0.1479   &     0.3474    &     0.1905      \\
\multicolumn{1}{c||}{FPMC}                            &     0.1220   &    0.1813     &  0.3810            &     0.1431   &      0.1838   & 0.3165         \\
\multicolumn{1}{c||}{GRU4Rec}                         &     0.1860   &     0.2550    &    0.4758          &   0.1632     &   0.2050      &   0.3417       \\
\multicolumn{1}{c||}{SASRec}                          &     0.1901   &     0.2612    &    0.4920          &   0.2009     &    0.2447     &   0.3874       \\
\multicolumn{1}{c||}{BERT4Rec}                        &  0.1819      &  0.2568       &   \underline{0.5061}           &   0.1313     &    0.1738     &   0.3135       \\ 
\hdashline
\multicolumn{1}{c||}{GRU4Rec$^+$}                        &    0.1880    &    0.2550     &    0.4758          &  0.1848      &  0.2294       &   0.3746       \\
\multicolumn{1}{c||}{SASRec$^+$}                         &    \underline{0.2022}    &    \underline{0.2710}     &    0.4970          &   0.2045     &  0.2488       &  0.3930        \\ 
\multicolumn{1}{c||}{FDSA}                            &     0.1913   &     0.2625    &     0.4984         &    0.2056    &    0.2522     &    0.4040      \\
\multicolumn{1}{c||}{MLP-Mixer$^+$}                       &  0.1987      &  0.2671       &    0.4920          &    \underline{0.2089}    &    \underline{0.2556}     &      \underline{0.4065}    \\
\multicolumn{1}{c||}{\textbf{MLP4Rec}}                         &    \textbf{0.2027}    & \textbf{0.2747}        &    \textbf{0.5118}          & \textbf{0.2139*}       &   \textbf{0.2654*}      &    \textbf{0.4326*}      \\
\arrayrulecolor{black}\hline
\end{tabular}
\\``\textbf{{\large *}}'' indicates the statistically significant improvements (i.e., two-sided t-test with $p<0.05$) over the best baseline.
\label{table:overall}
\vspace{-3mm}
\end{table*}

\subsection{Implementation Details}

The implementation of MLP4Rec and all baselines are based on RecBole's library~\cite{zhao2021recbole}, an open-source recommender system library, which allows us to test and compare all methods in a fair environment, and allows our results to be reproduced easily. 

We tune the hyperparameters based on original papers' recommendations. If original paper did not provide detailed hyper-parameters, we perform hyper-parameter tuning via cross-validation with Adam optimizer~\cite{kingma2014adam} and early stop strategy. Due to the limited space, please refer to the \textit{Section D of the Technical Appendix} for detailed hyper-parameter tuning ranges and results, as well as \textit{Section E of the technical appendix} for other detailed experiment settings.

\subsection{Performance Comparison}

We will compare our proposed methods against following baselines: PopRec, BPR ~\cite{rendle2012bpr}, FPMC ~\cite{rendle2010factorizing}, GRU4Rec ~\cite{hidasi2015session}, GRU4Rec$^+$~\cite{hidasi2016parallel}, SASRec and SASRec$^+$~\cite{kang2018self}, BERT4Rec ~\cite{sun2019bert4rec}, FDSA~\cite{zhang2019feature}, and MLP-Mixer$^+$~\cite{tolstikhin2021mlp}. Note that (1) superscript ``+'' means that we improve the original model, which takes the concatenation of embeddings of item ID and features as input, enabling fair comparison with MLP4Rec; (2) The details of baseline methods are available in the \textit{Section F of the Technical Appendix}. 

Table~\ref{table:overall} summarizes the comparison results, where models above the dashed line consider only item embeddings, while below models also involve item features. From Table~\ref{table:overall}, we can make the following general observations: (1) Starting from GRU4Rec, deep learning based methods exceed traditional methods such as BPR by a large margin, suggesting that in sequential recommendation, deep learning models are better at capturing sequential dependencies. (2) Models that can handle item features (e.g. SASRec$^+$, FDSA) usually outperform those who cannot (e.g. SASRec, BERT4Rec), indicating the importance of item features in sequential recommendations. (3) Improvement over the best baseline is more significant on the larger dataset ``Beauty''. More specifically, we can also observe that: (4) Compared to RNN-based models, self-attention models usually have better performances, which can be attributed to self-attention's stronger capabilities in capturing sequential patterns. (5) MLP-Mixer$^+$ can achieve comparable performance when compared with the SOTA methods such as SASRec and FDSA.
(6) MLP4Rec constantly outperforms all baselines including MLP-Mixer$^+$ with a significant margin, which suggests that tri-directional information fusion is an important improvement, which jointly captures sequential, cross-channel, cross-feature correlations, whereas SOTA baselines like SASRec or FDSA ignore the correlations between item features at item level, resulting in subpar performance.

\begin{table}[t]
\centering
\caption{Model complexity comparison on Beauty dataset. } 
\vspace{-3mm}
\begin{tabular}{cccc} 
\hline
 Model & Param & NDCG@10 & HR@10  \\ 
\hline
BERT4Rec & 2.3M & 0.1738 & 0.3135  \\
GRU4Rec$^+$ & 2.5M & 0.2294 & 0.3746  \\
SASRec$^+$ & 2.0M & 0.2488 & 0.3930  \\
FDSA & 3.2M & 0.2522 & 0.4040  \\
MLP-Mixer$^+$ & 1.8M & 0.2556 & 0.4065\\
\textbf{MLP4Rec} & \textbf{1.7M} & \textbf{0.2654} & \textbf{0.4326}  \\
\hline
\end{tabular}
\\\small{``Param" refers to the number of trainable model parameters.} 
\vspace{-3mm}
\label{table:complexity}
\end{table}

\subsection{Model Complexity} 
As shown in Table~\ref{table:complexity}, despite MLP4Rec's superior performance, it also surpasses baselines in terms of memory efficiency. Fewer model parameters not only make the MLP4Rec easier to train, but also reduce the risk of over-fitting ~\cite{lee2021moi}. It is also noteworthy that not only we have fewer parameters than self-attention models and RNN-based models, our method is also simpler than MLP-Mixer$^+$, which could be attributed to the removal of fully-connected embedding layer mentioned in Section F.2 of Technical Appendix.

\subsection{Parameters Analysis} 
\begin{figure}[t]
     \centering
    {\subfigure{\includegraphics[width=0.49\linewidth]{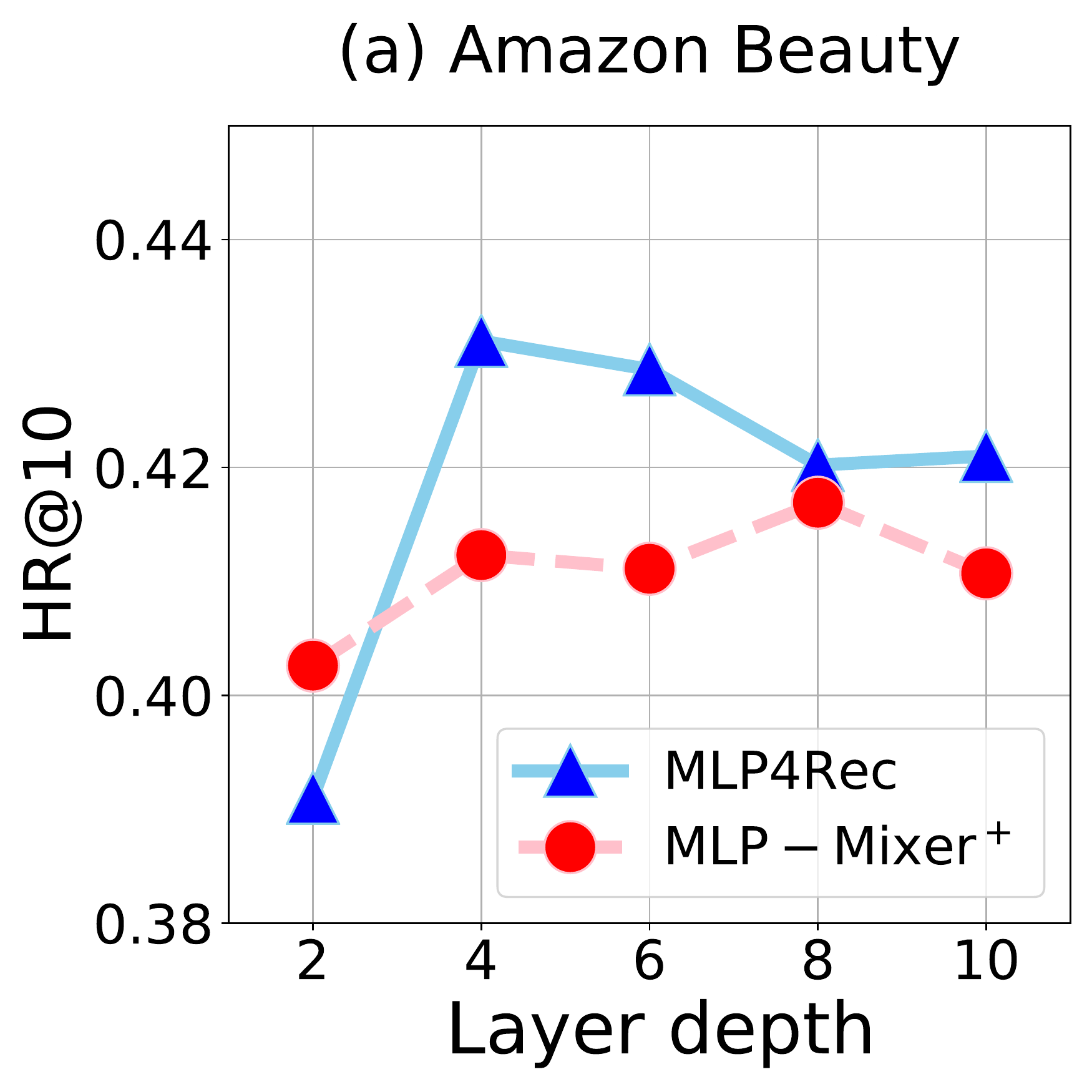}}}
    {\subfigure{\includegraphics[width=0.49\linewidth]{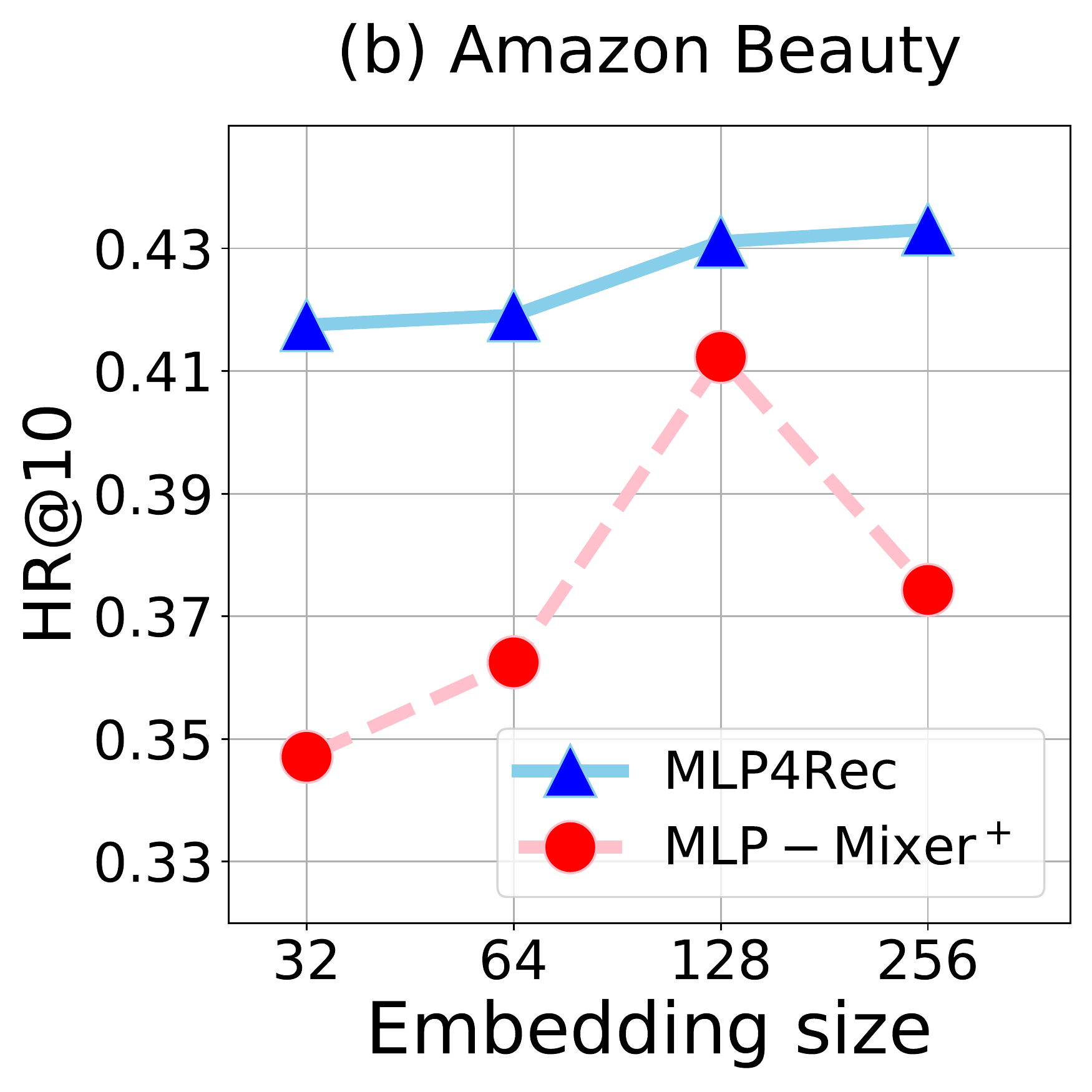}}}
    \vspace{-5mm}
    \caption{Influence of hyper parameters on performance}
    \label{fig:parameter}
    \vspace{-4.9mm}
\end{figure}

Figure~\ref{fig:parameter} shows the influence of layer depth and embedding size to MLP4Rec and MLP-Mixer$^+$. Generally, unlike MLP-Mixer's application in CV ~\cite{tolstikhin2021mlp}, our framework in SRS does not require a very deep network. In addition, compared to MLP-Mixer$^+$, the performance of MLP4Rec is more robust across a wide range of embedding sizes. A potential reason for this is that the tri-directional information communication allows latent representations to be shared on different embedding tables, thus a smaller embedding size does not significantly harm the representational capacity of the model. However, MLP-Mixer$^+$ needs to compress rich semantics from item features. Thus, small embedding sizes (e.g., $32$ and $64$) lead to sub-optimal performance due to their limited representational ability. In contrast, a large embedding size (e.g., $256$) results in over-fitting issue.

\subsection{Ablation study}
As shown in the previous subsections, MLP4Rec achieves better performance than MLP-Mixer$^+$ in both datasets across all metrics, and the only difference between their architecture is feature-mixer. Here, we investigate the necessity of a feature-mixer by answering two important questions: \textit{Q1: Can simpler alternatives achieve the same performance as the feature-mixer?} and \textit{Q2: What are the contribution of each respective modules in our proposed model?} To answer those questions, and further validate the importance of our proposed improvement, we design following alternatives to MLP4Rec and MLP-Mixer:
(1) {\bf MLP-Mixer} is the vanilla MLP-Mixer which does not include item features. 
(2) {\bf MLP4Rec-Linear} is the simplified MLP4Rec which replaces the MLP feature-mixer block with a simple linear layer. 
(3) {\bf MLP4Rec-Simple} is the simplified MLP4Rec which only performs feature mixing at the final layer instead of every layer.
(4) {\bf w/o Sequence-Mixer} is the simplified MLP4Rec without Sequence-Mixer module. 
(5) {\bf w/o Channel-Mixer} is the simplified MLP4Rec without Channel-Mixer module. 
(6) {\bf w/o Feature-Mixer} is the simplified MLP4Rec without Feature-Mixer module, which is equivalent to vanilla MLP-Mixer.

The performances of alternatives are displayed in Table~\ref{table:ablation}, where the upper-part of the table mainly addresses \textit{Q1} and lower-part of the table mainly addresses \textit{Q2}. From Table~\ref{table:ablation}, we can summarize that
(1) Without incorporating item features, MLP-Mixer has significantly worse performance, which confirmed the importance of introducing item features into sequential recommendations.
(2) For two simplified MLP4Rec versions, we can observe that MLP4Rec-Linear constantly outperforms MLP4Rec-Simple, which means that only communicating feature information at the last layer cannot fuse cross-feature correlation into the hidden representation sufficiently. 
(3) MLP4Rec outperforms MLP4Rec-Linear and MLP4Rec-Simple consistently over all metrics, attributing to its full tri-directional fusion by feature-mixer. 
(4) Without Sequence-Mixer, MLP4Rec's performance degenerate most significantly, indicating that Sequence-Mixer plays a vital part in sequential recommendation, and can successfully capture the sequential pattern.
(5) Without Channel-Mixer, MLP4Rec also suffers a significant decrease in performance. The most likely reason is that, without the Channel-Mixer, the respective dimension of item/feature embedding cannot communicate with one another, thus making the hidden representation lack cross-channel correlation, and resulting in worse performances. And since without Feature-Mixer, the performance of MLP4Rec essentially equivalent to vanilla MLP-Mixer, so the effect of removing Feature-Mixer can refer to summarization (1).

\section{Related Work}
\label{sec:relatedwork}

In this section, we review the related work from the literature of sequential recommendation systems and MLP-Mixer. 

\subsection{Sequential Recommendation Systems}


RNN-based models can handle complex sequential dependencies in sequential recommendations by compressing previous user-item interactions into a vector that summarizes that information, and then make the prediction of the next possible interaction~~\cite{hidasi2015session,hidasi2016parallel,donkers2017sequential,quadrana2017personalizing,yu2016dynamic}. For example, GRU4Rec~\cite{hidasi2015session} is one of the most representative RNN-based SRS, which implements gated recurrent unit to improve the modeling of long-term dependencies, however, even with GRU, RNN-based models still cannot perform very well on a long sequence. 

Recent years, (self-)attention methods ~\cite{bahdanau2014neural,vaswani2017attention,li2017neural} show SOTA performances in SRS. SASRec ~\cite{kang2018self} is one of the first to implement self-attention for SRS and obtains promising results, by stacking several self-attention blocks, SASRec is able to learn dependencies among items.
BERT4Rec~\cite{sun2019bert4rec} implements bi-directional self-attention blocks and a Cloze objective, which also shows promising results. FDSA uses self-attention on both item token and item features to gain more information for better prediction. 
Nevertheless, self-attention's drawbacks are just as significant, whose computational complexity is quadratic to the length of the input sequence and embedding size.

\begin{table}[t]
\centering
\caption{Ablation study comparison}
\vspace{-3mm}
\begin{tabular}{cccc} 
\hline
 Model & MRR@10 & NDCG@10 & HR@10  \\ 
\hline
MLP-Mixer & 0.1974 & 0.2401 & 0.3790  \\
MLP4Rec-Linear & 0.2100 & 0.2586 & 0.4165  \\
MLP4Rec-Simple & 0.1995 & 0.2500 & 0.4143  \\
MLP-Mixer$^+$ & 0.2089 & 0.2522 & 0.4040  \\
\hdashline
w/o Sequence-Mixer & 0.1771 & 0.2153 & 0.3396 \\
w/o Channel-Mixer & 0.1933 & 0.2440 & 0.4092 \\
w/o Feature-Mixer & 0.1974 & 0.2401 & 0.3790 \\
\textbf{MLP4Rec} & \textbf{0.2139} & \textbf{0.2654} & \textbf{0.4326}  \\
\hline
\end{tabular}
\label{table:ablation}
\vspace{-5mm}
\end{table}

\subsection{MLP-based Architectures}

Recent development in MLP architectures reveals high potential in computer vision~\cite{tolstikhin2021mlp,touvron2021resmlp,liu2021pay}. Among them, MLP-Mixer ~\cite{tolstikhin2021mlp} is a symbolic example of recent advances in MLP-based models. Despite its simplicity, its performance is comparable with SOTA methods. MLP-Mixer uses token-mixer and channel-mixer to separately learn the spatial and channel correlations, with linear computation complexity and simpler architectures, MLP-Mixer was reported to have comparable performance compared with SOTA methods.

MOI-Mixer ~\cite{lee2021moi} is the first work to investigate the possibility of implementing MLP-Mixer in the sequential recommendation. They propose a Multi-Order-Interaction layer to improve its performance, while this improvement shows competitive performance, it is still outperformed by methods based on self-attention. In addition, it cannot handle item features. Until the submission of this paper, the code of MOI-Mixer has not been released, so we do not consider it as a baseline method for comparison. 
\section{Conclusion}
\label{sec:conclusion}
In this paper, we proposed a simple but efficient architecture with only MLP blocks for sequential recommendations. This architecture leverages a novel way to coherently connects sequential, cross-channel and cross-feature correlations in users' historical interaction data to mine their preference. MLP4Rec shows superior performances against state-of-the-art methods with a significant margin on two commonly used benchmark datasets, validating that: 
(1) MLP4Rec offers a powerful alternative to current self-attention based methods; 
(2) Feature-mixer enables the proposed model to cope with heterogeneous features and capture their correlations.
In addition, MLP4Rec's simpler model architecture and much fewer model parameters enhance its scalability in large-scale practical recommender systems.

\section*{ACKNOWLEDGEMENT}
This research was partially supported by Start-up Grant (No.9610565) for the New Faculty of the City University of Hong Kong and the CCF-Tencent Open Fund.

\bibliographystyle{named}
\bibliography{ijcai22}

\appendix

\title{Appendix}

\section{Model Optimization Algorithm}
In this section, we introduce the optimization algorithm of our model through pseudo-code, the PyTorch implementation could be found in our code submission. As shown in Algorithm 1, the structure of MLP4Rec is extremely simple, and sequence-mixer, channel-mixer, feature-mixer only perform simple matrix multiplications, therefore maintaining linear-complexity the whole time.

\begin{algorithm}[t]
	\caption{\label{alg:DARTS} Optimization algorithm of MLP4Rec}
	\raggedright
	{\bf Input}: Users' historical interaction data $S$\\
	{\bf Output}: Well-trained model \(f\)
	\begin{algorithmic} [1]
	    \STATE Randomly initialize parameters of model \(f\)
	    \FOR {Epoch in 1,...,max epochs}
	    \FOR {Batch 1,...,batch number}
	    \STATE Sample training batch data $S_{batch}$ from $S$
	    \STATE Generate predictions from \(f\)($S$)
		\STATE Calculate loss $L$ based on Equ (4)
		\STATE Update model parameters of \(f\) via minimizing $L$
		\ENDFOR
		\IF{Converged}
		    \STATE return \(f\)
		\ENDIF
		\ENDFOR
		\STATE return \(f\)

	\end{algorithmic}
\end{algorithm}

\section{Complexity Analysis}
\textbf{\textit{Time Complexity Analysis}}. Since each block of MLP4Rec is simply a MLP with one hidden layer, their time complexity is simply two matrix multiplications, i.e. \(i \times r + r \times o\), where \(i\) is the number of input units, \(r\) is the number of hidden units, \(o\) is the number of output units. Since the hidden units is a constant that independent from the input or output units, and in MLP4Rec the number of input unit is equal to the output unit, then we can denote the overall complexity as: $O(s + C + K)$, which is linear to the maximum sequence length \(s\), embedding size \(C\) and feature number \(K\). Compared to the complexity of self-attention, which is $O(s^{2}C + C^{2}s)$, we can conclude that the theoretical upper bound of the time complexity of MLP4Rec is significantly lower.

\textbf{\textit{Space Complexity Analysis}}. The learnable parameters in MLP4Rec consist of the parameters from the main body (mixers), embedding layer and output layer. 
In this section we will mainly be analyzing the number of parameters from the main body of the model. For each MLP block, the number of trainable parameters is essentially 2 matrices, their combined size is: \(i \times r + r \times o\). In addition, for each feature MLP4Rec trains separate sequence-mixer and channel-mixer, so the total space complexity of MLP4Rec is $O(K(s + C) + K)$, it's clear that the major influence is \(K\), which is the number of features, however in reality \(K\) usually tend to be relatively small, especially when it could be reduced by feature selection. On the other hand, the space complexity of self-attention is $O(sC + C^2)$~\cite{kang2018self}, which is quadratic to the embedding size. In experiment part, we show that not only we keep a theoretical lower upper bound in space complexity, but in practice, we also achieved significantly lower number of parameters. 

\section{Datasets and Features}

In this section we give more in-depth introduction of the features in our datasets, to show the performance of our proposed method on datasets that have relatively more features and datasets that relatively have fewer features, we use Amazon Beauty and MovieLens respectively. We cast away the features that seem less relevant to the prediction empirically, such as ``sales type" for Beauty (which only has a single value ``Beauty" for all items). 

\begin{table}[t]
\centering
\caption{Feature details.}
\label{table:statistics}
\vspace{-3mm}
\begin{tabular}{c|c|c} 
\hline
Feature name & Dataset & Feature type  \\ 
\hline
\ class & MovieLens & Textual token sequence  \\
\ movie title & MovieLens & Textual token sequence\\
\ release year & MovieLens & Float token\\
\ categories & Beauty & Textual token sequence  \\
\ sales rank  & Beauty & Float token  \\
\ price & Beauty & Float token \\
\ brand & Beauty & Textual token\\
\hline
\end{tabular}
\vspace{-3mm}
\label{table:featuredetails}
\end{table}

In Table \ref{table:featuredetails}, we show the names of the features we used, the dataset they belonged to, and their respective types. Textual token sequence refers to a sequence of keywords that indicates items' categories such as [``Action",``Comedy",``Drama"], we handle these types of features by first using Word2vec ~\cite{mikolov2013efficient} to project each of these keywords into individual dense vector representations, and then use average pooling to obtain an embedding that contains the overall semantic. Float token usually refers to features that are expressed in numbers, we use an embedding layer that can be viewed as a linear transformation to project the float token into embedding. And a textual token refers to an individual or group of words, but different from textual token sequence, the word in a textual token does not have individual meaning, such as ``Louis Vuitton", where ``Louis" and ``Vuitton" does not represent anything unless together they form a unique token, and we project such token into embeddings.

\section{Hyper-parameter Tuning}

\subsection{Hyper-parameter search range}

Tuning hyper-parameter is arguably one of the most important tasks when we want to evaluate models' performances. In order to show our effort in tuning baseline models as best as we could while balancing limited computational resources, we list our hyper-parameter searching range and the result of our parameter tuning. For simplicity, we divide our baseline and proposed methods into following sub-categories: \begin{itemize}

\item \textbf{A} including self-attention-based methods, which are SASRec, SASRec$^+$, BERT4Rec, FDSA.

\item \textbf{M} including MLP-based models, which are MLP-Mixer and MLP4Rec. 

\item \textbf{O} including other methods, which are PopRec, BPR and FPMC. 

\item \textbf{R} including RNN-based models, which are GRU4Rec and GRU4Rec$^+$. 

\end{itemize}

If original paper did not provide a guideline for selecting hyper-parameters, we tune them empirically based on their search range shown in Table \ref{table:hyper}, for validation we encourage you to use the hyper-parameter tuning modules of RecBole. Note that for hidden size in Table~\ref{table:hyper}, $2$ and $4$ refer to the ratio of hidden layer to the input layer. 

\begin{table}[t]
\centering
\caption{Hyper-parameter search range}
\label{table:hyper}
\vspace{-3mm}
\begin{tabular}{c|c|c} 
\hline
Model types & Hyper parameters & Search range  \\ 
\hline
\ A, R, M, O & embedding size & [32,64,96,128]  \\
\ A, R, M & dropout & [0,0.2,0.4,0.5]  \\
\ A, R, M & weight decay & [0,1e-3,1e-4] \\
\ A, R, M & hidden size & [2,4] \\
\ A, R & layers & [2,4] \\
\ A & heads & [2,4,8] \\
\ M & layers & [4,6,8] \\
\hline
\end{tabular}
\vspace{-3mm}
\label{fig:statistics}
\end{table}

\subsection{Tuning results}

In this subsection, we listed out the best hyper-parameters combination from defined search range under our experimental settings. 

\begin{table*}
    \centering
	\caption{Best hyper-parameter results}
	\label{table:statistics}
	\begin{tabular}{@{}|c|c|c|c|c|c|c|c|@{}}
		\toprule[1pt]
		Model & Dataset & embedding size & hidden size & weight decay & dropout& layers & heads  \\ \midrule
		\ \multirow{2}{*}{PopRec} & MoiveLens & - & - & - & - & - & - \\
		\ & Beauty & - & - & - & - & - & - \\
		\ \multirow{2}{*}{BPR} & MoiveLens & 64 & - & - & - & - & -  \\
		\ & Beauty & 128 & - & - & - & - & - \\
		\ \multirow{2}{*}{FPMC} & MoiveLens & 96 & - & - & - & - & -  \\
        \ & Beauty & 128 & - & - & - & - & - \\
		\ \multirow{2}{*}{GUR4Rec} & MoiveLens & 128 & 2 & 0 & 0.4 & 1 & - \\
		\ & Beauty & 128 & 2 & 0 & 0.4 & 1 & - \\
		\ \multirow{2}{*}{SASRec} & MoiveLens & 128 & 4 & 0 & 0.2 & 2 & 8 \\
		\ & Beauty & 64 & 2 & 0 & 0.5 & 2 & 4 \\
		\ \multirow{2}{*}{BERT4Rec} & MoiveLens & 128 & 4 & 0 & 0.5 & 4 & 4 \\
		\ & Beauty & 128 & 4 & 0 & 0.5 & 4 & 8 \\
		\ \multirow{2}{*}{GUR4Rec$^+$} & MoiveLens & 128 & 2 & 0 & 0.4 & 1 & - \\
		\ & Beauty & 128 & 2 & 0 & 0.4 & 1 & - \\
		\ \multirow{2}{*}{SASRec$^+$} & MoiveLens & 128 & 4 & 0 & 0.5 & 2 & 8 \\
		\ & Beauty & 128 & 4 & 0 & 0.5 & 2 & 8 \\
		\ \multirow{2}{*}{FDSA} & MoiveLens & 128 & 4 & 0 & 0.5 & 2 & 8 \\
		\ & Beauty & 128 & 4 & 0 & 0.5 & 4 & 8 \\
		\ \multirow{2}{*}{MLP-Mixer$^+$} & MoiveLens & 128 & 4 & 0 & 0.2 & 4 & - \\
		\ & Beauty & 128 & 4 & 0 & 0.5 & 4 & - \\
		\ \multirow{2}{*}{MLP4Rec} & MoiveLens & 128 & 4 & 0 & 0.4 & 4 & - \\
		\ & Beauty & 128 & 4 & 0 & 0.5 & 4 & - \\ \bottomrule[1pt]
	\end{tabular}
	\vspace{-2.1mm}
\end{table*}

\section{Implementation Details}

In this section, we will cover the details about experiments that cannot cover in the main body of the paper. We use Adam optimizer for optimization, learning rate is set to be 0.0001 , \(\beta_1\) = 0.9, \(\beta_2\) = 0.999 throughout the experiment. The training batch size is fixed at 256. Also, since MLP cannot handle different item sequence lengths, we use padding to fill users whose interaction number is less than maximum sequence length, and use most recent interactions from users who have more interactions than maximum sequence length\cite{kang2018self}. For a fair comparison, we set the non-linear activation across all models as GELU ~\cite{hendrycks2016gaussian}. Also, for BERT4Rec, mask ratio is also an important influence, accoring tothe suggestion of the original paper, we set that to be 0.2 for MovieLens dataset and 0.6 for beauty dataset. 

\subsection{Connection to previous study}

We note that one of the major concerns is ``why are the baseline performances of this study differs from previous study?" Though we adapt suggested parameters from models' original paper, we cannot completely replicate their experiment's settings, since previous works don't have the consistency on their respective experiments, also, not all works release their full settings.

\section{Baseline Methods}

\subsection{Model list}

In this part we will introduce the baseline methods from the experiment section of the paper in more detailed manner. 

\begin{itemize}[leftmargin=*]

\item \textbf{PopRec} simply recommends most popular items to user and does not account for user interest or item relevance.

\item \textbf{BPR} ~\cite{rendle2012bpr} builds matrix factorization model from pair-wise loss function to learn from implicit feedback, it is a classical general recommender system.

\item \textbf{FPMC} ~\cite{rendle2010factorizing} combines Markov Chains and Matrix Factorization method to learn the sequential dependencies in user interaction history as well as users' general preferences.

\item \textbf{GRU4Rec} ~\cite{hidasi2015session} uses Gated Recurrent Unit to improve the performance of traditional RNN in sequential recommender system. 

\item \textbf{GRU4Rec$^+$} ~\cite{hidasi2016parallel} is an improved version of GRU4Rec, it leverages item features and parallel computation to improve the performance.

\item \textbf{SASRec} ~\cite{kang2018self} introduces self-attention block for sequential recommendation. 

\item \textbf{SASRec$^+$} is a naive improved version of SASRec, we append feature embedding of items together with token embedding for fair comparison with our proposed method, this implementation was reported to have competitive performance in several studies ~\cite{zhang2019feature,zhou2020s3}. 

\item \textbf{BERT4Rec} ~\cite{sun2019bert4rec} uses bidirectional self-attention and mask-language-modelling (Cloze objective) method to make sequential recommendation. 

\item \textbf{FDSA} ~\cite{zhang2019feature} applies separate self-attention block on token embedding and feature embedding to learn their correlations.

\item \textbf{MLP-Mixer$^+$} ~\cite{tolstikhin2021mlp} is our improved version of MLP-Mixer to make it adapt to sequential recommendation tasks with item explicit features. Specifically, it takes the concatenation of embeddings of item ID and features as input, enabling fair comparison with MLP4Rec. We maintained the overall structure of MLP-Mixer$^+$ and make sure the only structural difference between MLP4Rec and MLP-Mixer$^+$ is feature-mixer layer.

\end{itemize}

\subsection{Modification details of MLPMixer$^+$}

When items that users interacted with have no features but their IDs, MLP4Rec is equivalent to the MLP-Mixer. However, this is not a common scenario since in most of the recommendation settings, items have multiple features. Nevertheless, a vanilla MLP-Mixer can also leverage item features for recommendation, MLP-Mixer was designed for image classification, and we can view the RGB channel of images as item features. MLP-Mixer first divides a 3-dimensional image (\(height \times width \times channel\)) into non-overlapping patches, and then project an individual 3-dimensional patch into a one-dimensional embedding vector, stacking the embedding vectors of different patches, the embedding table is then formulated as input.

Such an approach might work when cross-feature (channel) information is highly homogeneous such as different RGB channels for the same pixel. However, the cross-feature information for an item could be extremely heterogeneous, for instance, the embedding of the textual description of an item and the embedding of the brand of an item, that information have extremely diverged semantics. Moreover, in reality an item could have a large quantity of features, all of which can be heterogeneous from each other, so simply projecting all of them into an embedding while preserving the underlying semantics or pattern of each feature could be challenging. Therefore, the key distinction between our model and MLP-Mixer is the proposed feature-mixer layer. 
\end{document}